\documentstyle[11pt,dunk2001_asp,twoside]{article}
\def\cf{{\it cf.}}
\def\Msun{{M$_\odot$}}
\def\eg{{\it e.g.}}
\def\etal{{\it et al.}}
\def\etc{{\it etc.}}
\def\ie{{\it i.e.}}
\def\ltsim{\setbox1=\hbox{$<$}
 \;\hbox to \wd1{\raise.4ex\hbox{$<$}\kern-\wd1\lower.5ex\hbox{$\sim$}}\;}

\def\edcomment#1{\iffalse\marginpar{\raggedright\sl#1\/}\else\relax\fi}
\reversemarginpar

\begin{document}
\title{Distinguishing Dark Matter from Modified Gravity}
\author{J. A. Sellwood and A. Kosowsky}
\affil{Department of Physics \& Astronomy, Rutgers University, \\ 
136 Frelinghuysen Road, Piscataway, NJ 08854-8019, USA}

\begin{abstract}
Dynamical mass discrepancies in galaxies have two possible explanations: the existence of large amounts of dark matter or the breakdown of Newtonian gravity.  True dark matter halos of galaxies could differ in several respects from apparent halos inferred from misinterpreted modified gravity, and we discuss two observational tests to probe these differences.  Modified gravity gives generic, geometrical predictions for the shapes of apparent halos, and also requires that the inferred excess mass strongly correlates with the visible baryonic mass.  Thus modified gravity is more easily falsified, but we find both hypotheses to be consistent with the meager data on galaxy and cluster halos; dark matter is not {\it required\/} by current data.  We strongly encourage the acquisition of more data, which could ultimately eliminate one or other of these possibilities.
\end{abstract}

\section{Introduction}
The application of Newton's laws to galaxies and galaxy clusters leads to much higher mass estimates than expected from the observed stars and gas.  This well-established discrepancy is most commonly interpreted as evidence for dark matter, but it could also indicate a breakdown of Newton's laws in a regime (large scales and low accelerations) where they admit to essentially no other test.  Dark matter and modified gravity are the only two logical possibilities for explaining the data.  The dark matter hypothesis has consistently been favored, due to the formal elegance of general relativity and because the cosmological relic abundance of a massive particle with a weak interaction cross-section naturally gives the right order-of-magnitude mean dark matter density (\eg\ Lee \& Weinberg 1977).

It is far from clear how a modified gravity theory might reproduce the recent the successes of dark matter cosmology on large scales.  But the remarkable phenomenological regularity of galaxy rotation curves and the substantial difficulties faced by $\Lambda$CDM on galactic scales (Sellwood \& Kosowsky 2000) provides some motivation for continuing to explore modified gravity as an alternative to dark matter.  Here we propose observational tests that might distinguish between the two hypotheses, focusing on tests in the local universe; we discuss cosmological tests elsewhere.

The tests are completely independent of any specific theory for modified gravity.  The best known example is the simple, {\it ad hoc\/} modification to the Newtonian force law in the small-acceleration regime, known as MOND (Milgrom 1983).  This proposal, which was laid on more secure dynamical foundations by Bekenstein \& Milgrom (1984), is clearly falsifiable but has survived numerous confrontations with observations.  [Many of the criticisms collected by Scott \etal\ (2001) had been refuted previously.]  Detailed work has shown that the MOND force law predicts the observed rotation curves from the visible matter alone with remarkable accuracy (\eg\ Begeman, Broeils \& Sanders 1991).  These and other data indicate that MOND represents the {\it effective\/} force law for spiral galaxies, as emphasized by McGaugh (1999), suggesting that MOND might represent the {\it actual\/} gravitational force law arising from some broader modified gravity theory.  Dark matter theories face many fine-tuning problems in reproducing this effective force law, and the remarkable quality of the rotation curve fits indicate that MOND is more than simply a statistical consequence of the dark matter clustering hierarchy (\cf\ Kaplinghat \& Turner 2001).

Here we address a simple question: what observational tests might distinguish the modified gravity and dark matter hypotheses?  Any simple theory of modified gravity would be ruled out (and thus the existence of dark matter confirmed) by:

\begin{itemize}
\item Laboratory detection of the dark matter particle;

\item Dark matter halos having axes misaligned with those of the visible matter or having shapes inconsistent with geometrical requirements of modified gravity;

\item Evidence that visible matter is not a good predictor of total mass.

\end{itemize}

\noindent The laboratory test is already being pursued vigorously; a non-detection would be indecisive, although stringent enough limits on the dark matter interaction cross section would remove the natural particle physics motivation for weakly-interacting dark matter.

Dark matter is less easy to falsify, because its predictions on galaxy scales are less definite than those of modified gravity and because the properties of dark matter can be modified in an {\it ad hoc\/} manner.  Conversely, we may be able to find evidence that clearly rules out any sensible modification of gravity and therefore requires dark matter, \eg\ through discovery of a mass distribution clearly misaligned with the light.

\section{Laboratory Detection}
If galaxy halos are filled with dark matter particles, the Milky Way at the position of the Sun must possess a large local excess of dark matter particles.  A laboratory detection of a suitable candidate particle with the appropriate flux and mass would confirm dark matter and exclude modified gravity. 

Neutrinos are now known to have mass and therefore constitute a {\it bona fide\/} dark matter component of the universe.  The latest result from SNO (Ahmad \etal\ 2001) suggests the sum of the mass eigenvalues of active neutrinos is $> 0.05 \;$eV, while corresponding upper limit is $8.4 \;$eV, from the beta decay of tritium (Bonn \etal\ 2001).  Ahmad \etal\ place the neutrino mass density, expressed as a fraction of the critical density, in the range $0.1\% \ltsim \Omega_\nu \ltsim 18\%$ -- possibly exceeding $\Omega_{\rm Baryon}$ but probably rather less.  It is unlikely that neutrinos constitute the $\Omega_{\rm DM} \sim 0.25$ inferred from cosmological data (\eg\ de Bernardis \etal\ 2001), especially since hot dark matter has been out of favor for some time (\eg\ Gerhard \& Spergel 1992; Cen \& Ostriker 1992).  Assuming standard weak interactions and thermal equilibrium at decoupling (\eg\ Kolb \& Turner 1994), the allowed mass range ensures that neutrinos will move too rapidly to collect in the halos of individual galaxies (Tremaine \& Gunn 1979), and they will be irrelevant to internal galaxy dynamics.  Only if their masses are at the heavy end of the allowed range would they collect in galaxy clusters and make a detectable contribution to the CMB power spectrum.

As yet, there is no convincing laboratory detection of a CDM particle.  Current limits, summarized by Martineau (2001), almost exclude the recent detection claimed by Bernabei \etal\ (2000).

\section{Halo Shapes}
Reliable measurements of the shapes of halos have the potential to distinguish dark matter from modified gravity because the two hypotheses make quite different predictions.  Present knowledge of the three-dimensional shapes of halos is unfortunately rather sketchy (see recent reviews by Sackett 1999 and Merrifield 2001), but is improving rapidly.  Large samples of planetary nebulae and globular clusters far out in halos (\cf\ Ford, this meeting) are one example where fresh data may soon be brought to bear.

\subsection{Predictions}
The predictions for halo shapes in a $\Lambda$CDM universe have most recently been summarized by Bullock (2001).  The typical halo flattening of 2:1 found in older work (Frenk \etal\ 1988; Dubinski \& Carlberg 1991; Warren \etal\ 1992) applies only to SCDM while $\Lambda$CDM halos are generally somewhat closer to spherical.  Bullock examines many dark matter halos in cosmological simulations and finds a mean $c/a\sim 0.70$ within $30\;h^{-1}\;$kpc and a scatter of $\pm 0.17$ about the mean, with a long tail skewed towards highly flattened objects.  The same halos become generally flatter further out, and the principal axes of the mass distribution can twist through as much as right angles between the most tightly-bound region and the outer halo.  Baryonic infall makes weakly triaxial halos more nearly axisymmetric in the disk plane and slightly flatter (\eg\ Dubinski 1994; Evrard \etal\ 1994).  In summary, galaxy halos are expected to be slightly oblate spheroids in their inner parts and, at least in a statistical sense, to become generally more triaxial and twisted at larger radii.

A warm (less massive) dark matter particle (Colombi, Dodelson \& Widrow 1996; Hogan 1999), or one that is self-interacting (Spergel \& Steinhardt 2000), have been proposed to ameliorate some of the better-known difficulties of $\Lambda$CDM.  In the same study, Bullock finds that halos in $\Lambda$WDM simulations are slightly closer to spherical.  Closely spherical halos are a firm prediction of SIDM (Dav\'e \etal\ 2001; Miralda-Escud\'e 2000); because the particles have an isotropic velocity distribution, the gravitational field of the disk causes very mild flattening.

If mass discrepancies are in fact produced by a modification to gravity, simple geometry dictates that the apparent mass distribution inferred using Newton's law would indicate: (1) an asymptotically spherical dark matter halo away from the disk; (2) a highly flattened halo in the inner parts near the disk; and (3) halo axes aligned with the light axes.  Milgrom (2001) has recently emphasized that gravitational potentials computed using the Bekenstein \& Milgrom (1984) modified Poisson equation would appear to a Newtonist as arising from halos having these properties.

We therefore conclude that modified gravity could be excluded if inner halos are not highly flattened or if outer halos have increasing non-axisymmetry towards larger radii or twists of the mass distribution away from the plane of the luminous matter.  On the other hand, halo shapes that fit with the modified gravity predictions would not explicitly rule out dark matter, although the discovery of many highly flattened inner halos would be out of line with current predictions.

\subsection{Halo Flattening}
Probes of the potential outside the principal plane are not easy to find or measure, making it hard to estimate the flattening of the halo.  Almost all reported estimates give a single value for the axis-ratio, $c/a$, of an oblate spheroid which obviously applies to the measured radius, but may not characterize the entire halo.  Moreover, Olling \& Merrifield (2000) note a worrisome tendency for different methods to yield systematically different results.  In particular, it seems that values of $c/a \ltsim 1$ which result from modelling warps of galaxy disks as uniformly precessing modes are unreliable, probably because the critical assumption of a long-lived mode is unlikely to hold.

One of the most detailed studies, which offers probably the most reliable single measurement, is by Sackett \etal\ (1994).  Using the observed velocities of material in the two almost perpendicular planes of the polar ring galaxy NGC~4650A, they conclude that the inner dark matter halo is as flat as $0.3 \ltsim c/a \ltsim 0.4$ in density.  Furthermore, it has already been verified that the observational data are consistent with the MOND prediction (Morishima \& Saio 1995), even though MOND does not have the extra free parameter of halo flattening.

X-ray halos around some massive early-type galaxies have also been used to probe the gravitational potential.  Buote \& Canizares (1997, 1998) reached some tentative conclusions on the basis of low resolution ROSAT data.  In a preliminary report of their recent observations of NGC~720 with CHANDRA, Buote \& Canizares (2001) show that their earlier data were contaminated with unresolved point sources, and the new data show significant, and radially decreasing, flattening, but rather less than originally claimed.  They also find evidence for a misalignment between the major axes of the light and of the gravitational potential well, which increases at large radii.  Modified gravity would be in serious trouble if the misalignment holds up in their further analysis and if the X-ray gas is demonstrated to be in hydrostatic equilibrium.

A similar analysis (Buote \& Canizares 1996, 2001) finds that galaxy clusters are flattened in their centers, but all become somewhat rounder at large radii.   The use of these data to test modified gravity is not straightforward, however, since the gas itself is no longer just a tracer, but becomes the dominant contributor to the mass.

Olling (1996) uses flaring of the gas layer beyond the edge of the optical disk to estimate that the inner halo of NGC~4244 is as flat as $c/a \sim 0.2_{-0.1}^{+0.3}$, if the gas velocity dispersion tensor is isotropic.  Milgrom (2001) has already shown that the MOND prediction for this galaxy is for a highly flattened apparent halo.  It is worth noting that a halo of dark matter particles that is flatter than about 3:1 could not survive, and would puff up through dynamical buckling instabilities (Merritt \& Sellwood 1994).

The Milky Way is the only galaxy for which shape estimates have been attempted at multiple radii.  Olling \& Merrifield (2000) use flaring of the gas layer to argue for $c/a \sim 0.8$, but reach this conclusion only by adopting galactic parameters ($V_0$ and $R_0$) that differ significantly from those recommended by the IAU.  As the standard values led them to conclude that the halo is prolate, it is unclear whether this method yields sensible results.  Halo stars are perhaps more reliable probes of the potential: van der Marel (2001) deduces $c/a > 0.4$ for the inner halo from the velocity ellipsoid of local halo stars.  Ibata \etal\ (2000) argue, on the basis of just 38 halo carbon stars, that the Saggitarius stream moves in a near-spherical halo at large radii.   Note that this extremely tentative evidence for a flattened inner halo which becomes more spherical further out is consistent with the modified gravity prediction, but will be subject to increasingly robust tests as these constraints improve.

\subsection{Axisymmetric Halos}
The tightness of the Tully-Fisher relation precludes strongly non-axisymmetric halos in the region which produces the measured velocity widths (Franx \& de Zeeuw 1992), but this constraint is generally thought to be consistent with CDM predictions (Sackett 1999) and is, of course, also expected in modified gravity.  Intrinsically non-axisymmetric {\it disks\/} have been found (\eg\ Rix \& Zaritsky 1995; Andersen \etal\ 2001), but again are not clear evidence of non-axisymmetry in the dark matter, since the luminous mass itself could well be responsible.

Galaxies with tracers outside the bright part offer sharper tests: Schoenmakers (1998) has analyzed extended HI disks around galaxies and notes that it is hard to disentangle any possible distortions due to a non-axisymmetric halo from the influence of warps and spirals in the baryonic matter.  He feels that the data probe the halo shape in two cases: he finds NGC~3198 to be closely axisymmetric at all radii outside the disk and the trend in the halo shape of NGC~2403 is towards becoming more axisymmetric further out.

The most stunning case by far is that of IC\,2006 (Franx, van Gorkom \& de Zeeuw 1994).  The shape and kinematics of the gas ring surrounding this early-type galaxy suggest that it is inclined as the galaxy and that departures from axisymmetry in the potential are $\ltsim 1\%$ at $R > 6 R_e$.

\subsection{Halo Alignment}
Kochanek (2001) brings strong lensing data to bear on the question of halo shapes.  From an analysis of 20 separate strong lenses by galaxies, and taking the large-scale shear into account, he concludes that the prinicipal axis of the lensing mass is unlikely to be misaligned by more than 10$^\circ$ from that of the light distribution.  He concludes that ``mass follows light,'' even though he estimates that more than half the lensing mass lies in a volume outside that containing the light.

Although light deflection may be different in modified gravity theories, symmetry considerations require certain geometric regularities in lensing regardless of the theory.  Thus we can very reasonably expect both the strong deflections and the weak-shear orientation to correlate with equipotential lines, regardless of the actual deflection law.  Without an explicit lensing law, we cannot use particular lensing patterns to measure masses, although they might be used as a direct probe of the deflection law (\eg\ Mortlock \& Turner 2001). 

\section{Light as a Predictor of Mass}
Modified gravity could also be excluded if light (stars and gas) were a poor predictor of dynamical mass.  In the dark matter hypothesis, the expected universal baryon fraction could, for many possible reasons, yield different fractions of mass in stars.  Even within halos of similar mass, different merging histories, ionizing fluxes, \etc, should lead to statistical variations in the luminous mass.  Such variations could not arise in the modified gravity picture, whereas they are an expected feature, described as {\it bias}, in CDM models: the early predictions (\eg\ White \etal\ 1987) were rather na\"\i ve, but it now seems that a scale-dependent bias is required to match the clustering observations (\eg\ Peacock \& Smith 2000; Hamilton \& Tegmark 2000).

\subsection{Light traces mass}
Kalnajs (1983), Kent (1986), Buchhorn (1992), Broeils \& Courteau (1997) and Palunas \& Williams (2000) have found for many hundreds of galaxies that the light distribution within a galaxy is an excellent predictor of the shape of the inner part of the rotation curve.  The success of Milgrom's formula for predicting the entire rotation curve from the observed baryonic content (\eg\ Begeman \etal\ 1991) indicates an extraordinarily strong coupling between the mass profiles of the luminous and dark matter within galaxies.

The Tully-Fisher relation is an empirical correlation between galaxy luminosities and velocity widths.  The velocity width clearly depends on the radial density profile, and is therefore related in some way to the mass.  The tightness of this correlation (\eg\ Sakai \etal\ 2000; Verheijen 2001) again betrays a close correlation between light and mass.

Bahcall, Lubin \& Dorman (1995) find a rather narrow trend in deduced M/L$_{\rm B}$ as a function of spatial scale from small galaxies to galaxy groups and clusters.  But it is the smallness of the scatter among similar objects in their plots which is more interesting here than the trend.  The spread about the trend has a full-width of a factor of $\sim 4$, but is not purely random: early-type galaxies generally lie on the high side while late-types are systematically low.  It is therefore likely that the spread could be considerably reduced by adopting luminosities in a redder color band.

Analysis of the massive 2dF redshift survey led Peacock \etal\ (2001) to the conclusion that the ``bias'' parameter is close to unity, \ie\ that there is little bias and light is a fair tracer of mass on large scales.  Weak lensing has been used in three recent noteworthy studies: Hoekstra \etal\ (2001) find that light traces mass on scales $0.15 \ltsim l \ltsim 3 \, h_{50}^{-1}\;$Mpc.  Their result, combined with that from Peacock \etal\ (2001), is in disagreement with the predicted scale-dependent bias of $\Lambda$CDM.  McKay \etal\ (2001) use SDSS data to estimate the mass enclosed within a circle $260 h^{-1}\;$kpc around a galaxy which they find correlates well with the luminosity of the galaxy in most color bands.  Wittman \etal\ (2000) report measurements to much larger radii in galaxy clusters than have previously been achieved by the weak lensing method and find that the mass profile is closely proportional to that of the light within the cluster.  All these results again indicate that light is an excellent predictor of mass.

\subsection{A Counter-Example?}
Modified gravity would effectively be ruled out, and dark matter confirmed, if weak lensing revealed large mass concentrations unassociated with any light.  One such case has, in fact, been claimed: Erben \etal\ (2000) find a ``robust'' lensing signal suggesting a previously unidentified dark mass concentration of perhaps $10^{14}\;$\Msun.  Little X-ray emission is detectable from the position of this apparent mass concentration and there is no evidence for a galaxy cluster despite a deep H-band search (Gray \etal\ 2001).  If the lensing signal does indeed arise from a dark cluster-mass object, it would rule out any simple modified gravity proposal.  We note that such an object would also be problematic for dark matter since it is hard to imagine how it could avoid collecting enough hot gas to be X-ray bright.  It is possible that the lensing signal is spurious because of intrinsic alignments among the background galaxies and more conclusive evidence is needed before modified gravity is excluded.

\section{Conclusions}
Modified gravity as an explanation for mass discrepancies is a highly testable hypothesis.  Observations that could exclude it include: (1) detection of a halo misaligned with the light or not having the predicted shape, or (2) evidence that light (stars and gas) is not a good predictor of mass.  Not only would such data exclude modified gravity, but would necessarily support the existence of dark matter.

We have reviewed the existing data and find no convincing evidence that {\it requires\/} dark matter.  In particular:

\begin{itemize}

\item The meager data on halo shapes summarized in \S3 suggest that some inner halos are flattened, some outer halos are nearly axisymmetric, and most are aligned with the light.  All three results, though from only a few galaxies, are in line with the predictions of modified gravity, and none fit comfortably with the expectations of dark matter.  Some predicted CDM halos can be as flat as these estimates, but fewer are expected in WDM; SIDM would seem to be excluded by these data.

\item The fact that light is generally such an excellent tracer of mass (\S4) is remarkable and expected in the modified gravity hypothesis.  The one possible detection of a dark matter concentration with no associated baryons would, if confirmed, be a fatal problem for any modified gravity theory, however.

\end{itemize}

\noindent Existing data are inconclusive and more are needed.  It is noteworthy, however, that the easily-falsifiable modified gravity hypothesis is not excluded immediately, if the doubtful case can be set aside.  We conclude that dark matter is not screaming out from the currently available data.

\acknowledgments We thank Moti Milgrom and Stacy McGaugh for comments on a draft of this paper.  This work was supported by NSF Grant AST-0098282 to JAS and NASA grant NAG5-10110.  AK is a Cotrell Scholar of the Research Corporation.

\section*{Discussion}

\noindent {\it van der Kruit:\,} I always felt that the most compelling argument against MOND was the fact that two galaxies with completely different light distributions should not have the same rotation curves.  NGC~7814 is almost entirely an $R^{1/4}$-law bulge and NGC~891 is, in its light distribution, almost entirely an exponential disk with only a small bulge contribution.  Yet, both galaxies have identical rotation curves (about 220 km/s from 1~kpc out to the optical boundaries at 15 to 20 kpc).  I first said this at the Princeton IAU Symposium and documented it further in the {\it The Milky Way as a Galaxy}.

\noindent {\it Sellwood:\,} I am afraid I do not find your argument at all compelling for the reason that neither of these galaxies is easy to model.  The rotation curve of NGC~7814 is not well measured, as there is little HI gas in this early-type galaxy and the extreme velocity envelope is not easily traced.  The heavy dust obscuration in the edge-on galaxy NGC~891 makes modelling this galaxy difficult also: your model for the light distribution indicates an unusually low central surface brightness and a total luminosity which places it more than magnitude fainter than the mean of a B-band Tully-Fisher relation.  Until we have a good measurement in the NIR, we cannot exclude the disk being brighter, with a shorter scale length, than in your model.  One impeccable counter-example would indeed be enough to falsify MOND, but both these galaxies leave too much room for argument to be conclusive.

\noindent {\it Carignan:\,} You showed the good fits of Begeman \& Sanders, but many other galaxies require a different value for $a_0$ in order to get a good MOND fit.  Doesn't that discredit MOND?

\noindent {\it Sellwood:\,} As was clear from the case of NGC~2841 (Begeman \etal\ 1991), a good MOND fit requires an accurate distance because the acceleration is $V^2/R$, which is obviously distance-dependent.  If a wrong distance is adopted for a galaxy, a MOND-type fit will require a different value of $a_0$.  In cases of apparent discrepancy, it would be interesting to note what distance would be needed to obtain a reasonable MOND fit with the standard $a_0$ and whether such a revised distance would be out of the question.

\noindent {\it Quinn:\,} With large 8m class imaging, it is now possible to detect a galaxy-galaxy lensing signal from individual edge-on spirals.  This may allow us to directly measure halo shapes.

\noindent {\it Sellwood:\,} This is indeed another attractive method for addressing the issues raised here.  Small numbers of background galaxies, which might also have intrinsically correlated orientations, may mean that data from individual foreground galaxies are inconclusive.  But intrinsic halo shapes should be discernible with high confidence when data from many galaxies are combined.

\end{document}